\documentclass[aps,prb,reprint,twocolumn,superscriptaddress,amsmath,amssymb]{revtex4-1}
\usepackage{amsmath,amssymb}
\usepackage{graphicx}
\usepackage[tight]{subfigure}
\usepackage{longtable}
\usepackage{textcomp}
\usepackage{threeparttablex}
\begin{document}
%%%%%%%%%%%%%%%%%%%%%%%%%%%%%%%%%%TITLE%%%%%%%%%%%%%%%%%%%%%%%%%%%%%%%%%%%%%%%%%%%%%
\title{Ab-initio study of the relation between electric polarization and electric field gradients in ferroelectrics}
\author{J. N. Gon\c{c}alves}
\email{joaonsg@ua.pt}
\affiliation{Departamento de F\'isica and CICECO, Universidade de Aveiro, 3810-193 Aveiro, Portugal}
\author{A. Stroppa}
\affiliation{CNR-SPIN, 67100 L'Aquila, Italy}
\author{J. G. Correia}
\affiliation{Instituto Tecnol\'ogico e Nuclear, UFA, 2686-953 Sacav\'em, Portugal}
\author{T. Butz}
\affiliation{Fakult\"at f\"ur Physik und Geowissenschaften, Institut f\"ur Experimentelle Physik II, Universit\"at
Leipzig, Linn\'estrasse 5, 04103 Leipzig, Germany}
\author{S. Picozzi}
\affiliation{CNR-SPIN, 67100 L'Aquila, Italy}
\author{A. S. Fenta}
\affiliation{Departamento de F\'isica and CICECO, Universidade de Aveiro, 3810-193 Aveiro, Portugal}
\author{V. S. Amaral}
\affiliation{Departamento de F\'isica and CICECO, Universidade de Aveiro, 3810-193 Aveiro, Portugal}
\date{27 February 2012}
%%%%%%%%%%%%%%%%%%%%%%%%%%%%%%%%%%ABSTRACT%%%%%%%%%%%%%%%%%%%%%%%%%%%%%%%%%%%%%%%%%%
\begin{abstract}
The hyperfine interaction between the
 quadrupole moment of atomic nuclei and the electric 
field gradient (EFG) provides information on
 the electronic charge distribution close to a given atomic site. 
In ferroelectric materials, the loss of inversion symmetry of the 
electronic charge distribution is necessary 
for the appearance of the electric polarization.
We present first-principles density functional theory 
calculations of  ferroelectrics such as BaTiO$_3$, 
KNbO$_3$, PbTiO$_3$ and other oxides with perovskite structures, 
by focusing on both EFG tensors and polarization. 
We analyze the EFG tensor properties such as orientation and correlation
 between components and their relation with electric polarization. 
This work supports previous studies of ferroelectric materials 
 where a relation between EFG tensors and polarization 
was observed, which may be exploited to study the ferroelectric order 
when standard techniques to measure polarization are not easily applied.
\end{abstract}

\maketitle
%%%%%%%%%%%%%%%%%%%%%%%%%%%%%%%%%%%%%%%%%%%%%%%%%%%%%%%%%%%%%%%%%%%%%%%%%%%%%%%%
%%%%%%%%%%%%%%%%%%%%%%%%%%%%%%%%%%%%%%%%%%%%%%%%%%%%%%%%%%%%%%%%%%%%%%%%%%%%%%%%}
\section{Introduction}

There is great interest in ferroelectric/multiferroic materials 
nowadays due to their potential application in a plethora of subjects, 
ranging from high density memories to 
magnetoelectric sensors.~\cite{*[{}] [{ and references therein.}]
 Khomskii2009,Cheong2007,Bersuker2012} The complexity of electronic phenomena 
at the nanoscale  makes them a hot research topic and a fertile
 ground for new experimental 
techniques, which are able to probe point-like,
 atomic-scale properties. To this aim, the use of local probes as in 
hyperfine interactions techniques, such as  
M\"ossbauer effect spectroscopy, perturbed angular correlation 
(PAC) spectroscopy, nuclear magnetic resonance (NMR), and nuclear
 quadrupole resonance (NQR),
give access to atomic scale information 
of the electronic charge density~\cite{Schatz1996}
 through the measure of the electric field gradients 
(EFGs), thus probing the phenomenology of materials at the nanoscale. 

 From a theoretical point of view,  advances in
 modern density functional theory   have 
made the calculation of the spontaneous electric polarization $P$
 a routine calculation in an \textit{ab initio} framework.~\cite{Rabe2007,King-Smith1993,Resta1994} 
Recently, the calculation of EFGs has become also   
  possible from 
first-principles.~\cite{Blaha1985,Blaha1987,Blaha1988,Blaha1989,Ambrosch-Draxl1989,Ambrosch-Draxl1991,Blaha1992} 
However, to the best of our knowledge, 
 a theoretical study based on \emph{ab-initio}
 density functional methods  aiming to  investigate 
 both the  $P$ and EFGs in a ferroelectric material 
is still missing in the current literature despite the fact 
that a linear correlation~\cite{Oja1969,Fitzgerald1970,Fitzgerald1973}
between  $P$ and EFG values at a given site was shown a long time ago. This ``correlation'' 
would suggest that 
  information obtained  through the measurements 
of EFG tensors can provide 
indirect access  to the polarization as well. 
If true, one could study a macroscopic property of the crystal, 
such as $P$ by using local probes. In this work, we want to explore 
such a possibility.  

A nucleus with a nonspherical nuclear charge distribution possesses an electric quadrupole moment 
which leads to a hyperfine splitting for a nuclear spin $I\ge 1$ if subjected to an EFG.
The hyperfine techniques previously mentioned can measure the 
quadrupole coupling constant, which is the interaction between
 the nuclear quadrupole moment and the EFG. The EFG, in turn, 
arises due to the Coulomb potential at the nucleus, and its measurement is 
sensitive to the surrounding electronic charge density. 
More precisely, it is defined as the symmetric traceless second-rank 
tensor of second derivatives of the Coulomb potential with respect to the spatial 
coordinates, $V_{ij}=\partial^2 V/(\partial x_i \partial x_j)$,
 at the nuclear position. 
%Assuming non-overlapping electronic and nuclear charges it is traceless. 
% Provided the electronic charge density is obtained, the Poisson equation is solved
% to obtain the potential, and the EFG is then easily calculated in spherical coordinates
% from the spherical components of the potential as $V_{2}^m=\lim_{r\to0}\frac{1}{r^2}V_{2}^{m}(r)$. 
In the principal axis coordinate system, the tensor is 
diagonal and its elements are usually ordered by the convention
 $|V_{zz}| \ge |V_{yy}| \ge |V_{xx}|$. Usually $V_{zz}$ and 
the asymmetry parameter $\eta = (V_{xx} - V_{yy})/V_{zz}$ 
are used in the analysis of measurements. We recall that the EFG 
is site dependent, and its principal 
axes ($\tilde{x}$,$\tilde{y}$,$\tilde{z}$) 
may not be the same at every site, although they
 are usually along symmetry axes of the crystal.  
 EFG studies are found in various types of materials, for example, intermetallics,~\cite{Haarmann2011} metal complexes,~\cite{Bjornsson2010}
magnetic,~\cite{Asadabadi2007} or multiferroic compounds.~\cite{Lopes2011}

Previous studies have shown  that in some 
ferroelectric materials $P$ follows a temperature
 dependence which can be related to the EFG at specific 
atomic sites. For instance, NMR using $^{23}$Na in Rochelle salts [NaK(tartrate)$\cdot$4H$_2$O]  
showed that $P$ and the EFG are linearly  
related.~\cite{Oja1969,Fitzgerald1970,Fitzgerald1973} 
In 1978 Yeshurun suggested~\cite{Yeshurun1978} that the EFG 
due to static displacements  should be proportional to $P^2$
 in perovskite crystals using an empirical model for interpreting  
previous $^{57}$Fe M\"ossbauer 
measurements in BaTiO$_3$.~\cite{Bhide1966}
Dynamical aspects were also considered by  
relating the EFG to  the electric susceptibility, 
and it was found that the EFG should have a 
critical behavior when approaching $T_C$.~\cite{Yeshurun1978}
This peculiar feature was  recently used in 
the identification of ferroelectricity with
 EFG measurements.~\cite{Lopes2008}
 In this work, Pr$_{1-x}$Ca$_x$MnO$_3$ was 
studied with the measurement of the EFG at $^{111m}$Cd 
probes implanted into the sample. 
An abrupt change was found in a short temperature interval.
 This was associated to the onset of ferroelectricity, 
since the EFG should be dominated by a
 contribution proportional to the electric
 susceptibility at the transition, i.e.,  
with its critical behavior.~\cite{Yeshurun1978} In the same work,~\cite{Lopes2008}
it was also suggested that the temperature dependence of EFG tensors 
can give information on the onset of charge or orbital ordering. 
In Ref.~\onlinecite{Dening1980} it was argued that 
the static part of $V_{zz}$ should have the following 
behavior with respect to $P$: either it is proportional
 to $P^2$ in sites which have inversion symmetry in the 
paraelectric structure, or it is proportional to $P$,
 like in Rochelle salts.
The quadratic relation was supported by experiments
in NaNO$_2$,~\cite{Dening1980} in PbHfO$_3$ not too close to 
$T_C$ (where critical behavior is found), by PAC measurements,
% This technique measures the angular correlation of a 
% $\gamma$-$\gamma$ decay  cascade, emmited by probe
% isotopes introduced in the sample. The resulting anisotropy correlation 
% function has a perturbation during the lifetime of the intermediate 
% state due to the hyperfine interactions.
~\cite{Yeshurun1979,Schatz1996} and by NMR measurements in BaTiO$_3$.~\cite{Kanert1994}  
It is therefore clear from the current literature 
that $P$ and the EFG tensor are closely related quantities. 

Our study aims to explore  this relationship 
by calculating both $P$ and EFG for simple ferroelectric materials
and studying a possible correlation between these quantities. 
Some of the previously mentioned results are obtained by impurity probes in the 
host materials. 
Here, we shall limit our studies to systems 
where the probes are natural constituents of the materials.

This work is organized as follows. 
In Sec.~\ref{ComputDetails} we discuss 
the computational details. In Sec.~\ref{P_and_EFG} 
we present the results and discuss the relationship 
between $P$ and EFGs for simple tetragonal  
or orthorhombic (Sec.~\ref{orthorhombic}) systems.
We analyze the possible
 linear correlations between EFG tensor components 
in Sec.~\ref{czjzek}. A study of the   variation of $V_{zz}(P)$ with the atomic 
numbers of different materials is shown in section~\ref{Arelation}. Finally, in Sec.~\ref{Conclusions} we draw our 
conclusions.

\section{Technical details}\label{ComputDetails}

We have considered a series of simple $AB$O$_{3}$ type perovskite 
compounds.~\cite{King-Smith1994} For 
BaTiO$_3$, PbTiO$_3$ and KNbO$_3$ we have considered
 the tetragonal experimental structures as references. 
We also considered other perovskite-related  compounds,
such as BaZrO$_3$, CaTiO$_3$, PbZrO$_3$, SrTiO$_3$, 
NaNbO$_3$ and LiNbO$_3$  by considering a pseudo-cubic phase
 at the  experimental lattice constants of the cubic paraelectric phase.
The ferroelectric distortion was mimicked by a polar displacement 
of the atoms.
The experimental displacements in tetragonal BaTiO$_3$ are $z_{Ti}= 0.0203$, 
$z_{O1}= -0.0258$, and $z_{O2}= -0.0123$, in fractional 
 coordinates.~\cite{Kwei1993}  
 We calculate the EFG as a function of $\lambda$, which represents the fraction of the 
displacements ($z_{Ti}$, $z_{O1}$, $z_{O2}$) mentioned above. 
Therefore $\lambda=1$ corresponds to the equilibrium (experimental) structure. 
Values of $0 \le \lambda \le 1.2$ are used, i.e., $\lambda = 0$ corresponds to undistorted, $\lambda = 1$ 
corresponds to the experimental equilibrium distortion (at 280 K) 
and $\lambda = 1.2$ corresponds to 20\% additional distortion.
For the compounds where the cubic structure is used, 
a ferroelectric state is considered
 using the same fractional  distortions as in BaTiO$_3$. Although 
these states may not be observed in normal conditions, this 
allows us to study the possible correlation of polarization and 
EFG in different systems or as a function of strain. 
 \begin{table}\centering
\caption{\label{tab:latt} Lattice constants (in \AA{}) and atomic 
distortions used in the calculations of the perovskite compounds. 
$\delta_A$, $\delta_{B}$, $\delta_{O1}$, $\delta_{O2}$ 
correspond to the fractional distortions in the $c$ direction, 
with respect to the ideal positions, of the inequivalent sites. 
Sites $A$ are at the corners of the unit cell,
 $B$ is the transition metal inside an octahedron 
formed by apical O1 and equatorial O2 oxygen atoms.} 
\begin{threeparttable}
\begin{tabular}{lcccccc}
\hline\hline
Compound & $a$ & $c$ &$\delta_{A}$&$\delta_{B}$&$\delta_{O1}$&$\delta_{O2}$ \\ 
\hline
BaTiO$_3$\tnote{a} & 4.00 & 4.03 &&0.02 &-0.03&-0.01\\
KNbO$_3$~\tnote{b,c} & 4.00 & 4.06 &0.02&&0.04 &0.04 \\
PbTiO$_3$~\tnote{d,e} & 3.90  & 4.16&&0.04 &0.11&0.12\\
BaZrO$_3$\tnote{f} & 4.19  &  &&&&\\
CaTiO$_3$\tnote{f} & 3.83  &  &&&&\\
NaNbO$_3$\tnote{f} & 3.94  &  &&&&\\
PbZrO$_3$\tnote{f} & 4.13  &  &&&&\\
SrTiO$_3$\tnote{f} & 3.91  &  &&&&\\
LiNbO$_3$\tnote{g} & 4.00 &  &&&&\\
\hline\hline
\end{tabular}
\begin{tablenotes}
\item[a] Reference~\onlinecite{Kwei1993}.
\item[b] Reference~\onlinecite{Hewat1973}.
\item[c] Reference~\onlinecite{Shirane1954}.
\item[d] Reference~\onlinecite{Mabud1979}.
\item[e] Reference~\onlinecite{Nelmes1985}.
\item[f] Cubic lattices, experimental values (Ref.~\cite{King-Smith1994}).
\item[g] Theoretical lattice parameter, 
found by the volume optimization in the cubic phase, 
with atoms at the ideal positions.
\end{tablenotes}
\end{threeparttable} 
\end{table}

For the density functional theory calculations we used the projector-augmented-wave (PAW) method,~\cite{Blochl1994} 
as implemented in the \emph{Vienna ab-initio simulation package} (\textsc{VASP}),~\cite{Kresse1996} with 
the generalized gradient approximation (GGA)-Perdew, Burke and Ernzerhof (PBE) functional.~\cite{Perdew1996} 
% The electrons treated as valence states are: $1s^22s^1$, $2s^22p^4$, $2p^63s^1$, $3p^64s^1$, 
% $3p^64s^2$, $3p^63d^24s^2$, $4s^24p^65s^2$, $4s^24p^65s^24d^2$, $4p^64d^45s^1$, $5s^25p^66s^2$, $5d^{10}6s^26p^2$ for Li, 
% O, Na, K, Ca, Ti, Sr, Zr, Nb, Ba, and Pb respectively. 
% For BaTiO$_3$, a test with a slightly different atomic configuration, without Ti $3p$ electrons as valence, showed 
% negligible differences in $P$ and EFG. 
We used a $\Gamma$-centered Monkhorst-Pack~$7\times7\times7$ $k$-points grid, and an 
energy cutoff of 400 eV. The polarization is calculated with the Berry phase 
approach.~\cite{King-Smith1993,Resta1994} The EFG is calculated 
at the atomic sites $A$, $B$, apical O1, and equatorial O2 sites.

For the case of BaTiO$_3$ we also performed calculations with the L/APW+lo 
method, implemented in the \textsc{WIEN2k} code.~\cite{WIEN2k} The 
basis set was limited by $RK_{max}=8$, where $RK_{max}$ stands for the product of the smallest atomic 
sphere radius $R_{mt}$ times the largest $K$-vector $K_{max}$ of the plane wave 
expansion of the wave function, and a $6\times6\times6$ $k$-points grid was used.
%%%%%%%%%%%%%%%%%%%%%%%%%%%%%%%%%%%%%%%%%%%%%%%%%%%%%%%%%%%%%%%%%%%%%%%%%%%%%%%%
%%%%%%%%%%%%%%%%%%%%%%%%%%%%%%%%%%%%%%%%%%%%%%%%%%%%%%%%%%%%%%%%%%%%%%%%%%%%%%%%
%%%%%%%%%%%%%%%%%%%%%%%%%%%%%%%%%%%%%%%%%%%%%%%%%%%%%%%%%%%%%%%%%%%%%%%%%%%%%%%%
%%%%%%%%%%%%%%%%%%%%%%%%%%%%%%%%%%%%%%%%%%%%%%%%%%%%%%%%%%%%%%%%%%%%%%%%%%%%%%%%
\section{Results and discussion}\label{Results}

\subsection{Relation between EFGs and $P$}\label{P_and_EFG}
The lattice constants  are shown in table~\ref{tab:latt},
along with the experimental atomic distortions in
the ferroelectric compounds BaTiO$_{3}$, KNbO$_{3}$, and PbTiO$_{3}$. 
For the other cases, we have considered the fractional distortion of BaTiO$_3$.

\begin{figure}[tbp]
\begin{center} 
\subfigure[]{\label{ene_BaTi}\includegraphics[width=0.6\linewidth]{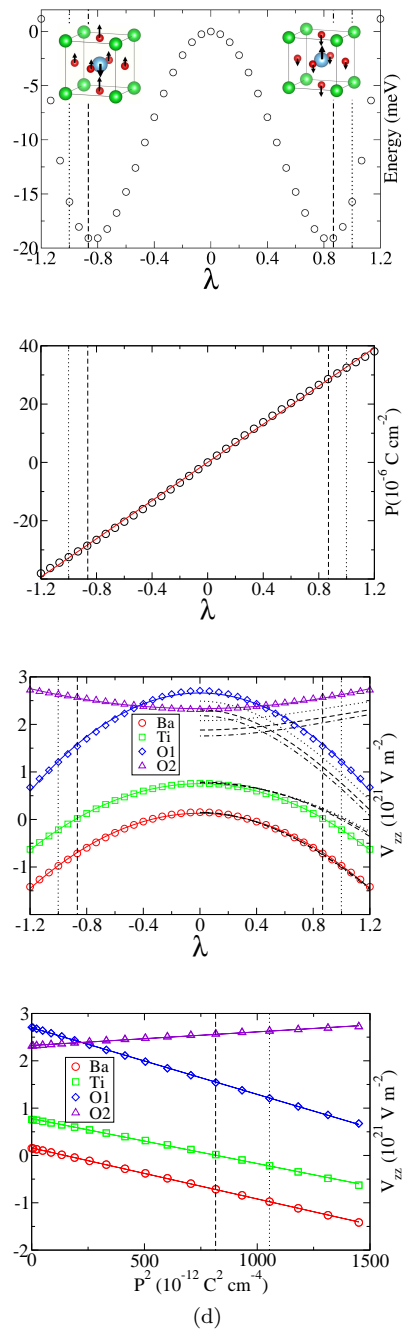}}
\subfigure[]{\label{pol_BaTi}\includegraphics[width=0.6\linewidth]{fig1b.eps}}
\subfigure[]{\label{disVzz_BaTi}\includegraphics[width=0.6\linewidth]{fig1c.eps}}
\subfigure[]{\label{pol2Vzzall_BaTi}\includegraphics[width=0.6\linewidth]{fig1d.eps}}
\caption{(Color online) BaTiO$_3$ in the tetragonal phase. (a) Energy (meV) as 
a function of distortion, varied in equal intervals from 0 to 1.2. 
The direction of the displacements 
is shown at each side in the plot. (b) $P$ ($\mu$C\,cm$^{-2}$) as a function of distortion. The vertical dashed 
and dotted lines correspond to the calculated energy minimum and 
experimental distortion, respectively. (c) $V_{zz}$ ($10^{21}$ V$\,m^{-2}$) as a 
function of distortion (dotted, dashed and dot-dashed lines are L/APW+lo calculations, see text). 
(d) $V_{zz}$ as a function of $P^2$ ($10^{-12}$ C$^2\,$cm$^{-4}$). 
Full lines are fits to the data.}
\end{center}
\end{figure}

Let us start by considering the case of  BaTiO$_3$. 
The variation of the total  energy with the  atomic distortions 
is shown in Fig.~\ref{ene_BaTi}, with the displacements considered 
along the polar axis $z$ in both directions. The curve shows the expected double well 
profile. The energy minimum is reached for 85\% of the experimental distortion: the slight 
discrepancy is probably due to the approximation used here for the exchange-correlation functional. 
Given that both the experimental/theoretical displacements are very small, this discrepancy is reasonable.~\cite{Note1}  
In all the following cases, the vertical dashed and dotted lines correspond to the theoretical energy minimum 
and the experimental distortions, respectively. The stable  state  has an energy lower than the undistorted 
one by $20$ meV, which compares well with previous calculations.  

Figure~\ref{pol_BaTi} presents $P$ as a function of
 the ferroelectric distortion.
It can be seen that $P$ is
 approximately a linear function of the distortion. 
This is not unexpected 
since we are considering displacive 
type ferroelectrics, as already discussed in    
previous works.~\cite{Resta1993}  The calculated $P$ at the 
experimental distortions is $28.6$ $\mu$C\,cm$^{-2}$, 
according to what can be seen in Fig.~\ref{pol_BaTi}, while the 
experimental one~\cite{Wieder1955} is $27$ $\mu$C\,cm$^{-2}$. 

In Fig.~\ref{disVzz_BaTi} we show  the EFG component $V_{zz}$ as
a function of distortion for all sites.~\cite{Note2} It is evident that  
$V_{zz}$ shows a quadratic dependence upon the distortion. 
The sign~\cite{Koch2009} and magnitude of the values obtained are consistent 
with previous calculations.~\cite{Alonso2004,Koch2009} 
Measurements~\cite{Blinc2008} obtained at the O sites by NMR are in agreement for the equatorial site with $2.56$, 
but at the apical site the experimental value is $2.06$ while our calculated value is smaller, $1.2$ (all in units of $10^{21}$ V\,m$^{-2}$).

Furthermore, we investigated how the EFG results depend on the 
choice of method or functional [see fig.~\ref{disVzz_BaTi}]. We have performed calculations using the 
L/APW+lo implementation of DFT, with the \textsc{WIEN2k} code. Three 
different functionals were used, which are represented in the plot 
by dotted, dashed,  and dot-dashed lines, corresponding to the 
GGA-PBE,~\cite{Perdew1996} GGA-Wu and Cohen (WC),~\cite{Wu2006} and local-density approximation (LDA)~\cite{Perdew1992} exchange-correlation functionals. 
The variation of EFG with distortion is the same, except for a small 
difference at Ti between the two implementations. There is 
also a shift in values at the O atoms, 
with a maximum difference of $0.5\times10^{21}$ V\,m$^{-2}$, when 
comparing LDA L/APW+lo and PBE PAW calculations, while maintaining 
the same variation with distortion. These differences are reasonable and  
the main feature, quadratic variation of $V_{zz}$ with distortion, remains the same. 
All the following results are taken from the PAW calculations.

In order to get rid of the distortion parameter, in Fig.~\ref{pol2Vzzall_BaTi} we plot the $V_{zz}$ EFG component 
as a function of $P^2$. In this case, a fit of the data 
clearly shows a linear dependence. 

\begin{figure}[tbp]
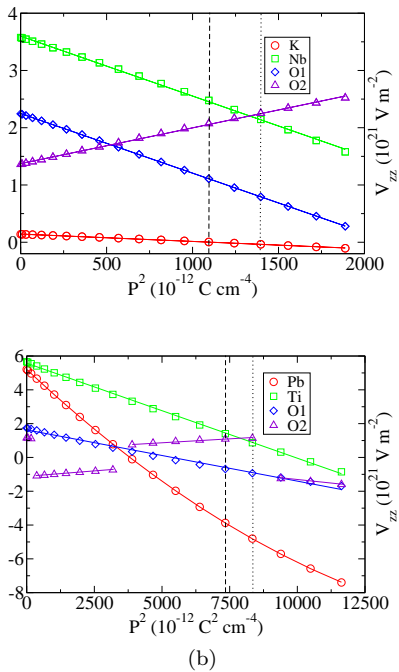

\begin{center} 
\subfigure[]{\label{pol2Vzzall_KNb}\includegraphics[width=0.6\linewidth]{fig2a.eps}}
\subfigure[]{\label{pol2Vzzall_PbTi}\includegraphics[width=0.6\linewidth]{fig2b.eps}}
\caption{(Color online) $V_{zz}$ at each site as a function of 
$P^2$ ($10^{-12}$ C$^2$\,cm$^{-4}$). The lines are fits to the $V_{zz}(P^2)$ data. (a) KNbO$_3$, (b) PbTiO$_3$,
 in the tetragonal phases.}
\end{center}
\end{figure}

For KNbO$_3$ the trends are similar, 
and a  linear  relation is also obtained in Fig.~\ref{pol2Vzzall_KNb}. 
In this case, the calculated value of $P$ at the experimental 
distortion is $37.4$~$\mu$ C\,cm$^{-2}$, consistent with the measured value.~\cite{Resta1993} 
The calculated $V_{zz}$ at Nb, $2.6\times 10^{21}$ V\,m$^{-2}$ is also 
consistent with $|V_{zz}|=2.7\times10^{21}$ V\,m$^{-2}$ obtained by an NMR experiment 
at $220$ C (Ref.~\onlinecite{Hewitt1961}) [considering $Q_{^{93}Nb}=-0.37 b$ (Ref.~\onlinecite{Stone2005})].

For the case of PbTiO$_3$, presented in Fig.~\ref{pol2Vzzall_PbTi}, much 
larger values of polarization are obtained. The polarization is 
$85$ ($91$) at the calculated (experimental) distortion, higher than 
the measured~\cite{Hewat1973} $75$~$\mu$ C\,cm$^{-2}$ at room temperature. Notably 
large variations of the EFG in the range of distortions are seen at the Pb and Ti sites. 
The obtained values of $V_{zz}$ are in reasonable agreement with Ti 
NMR experiments (Table 3 of Ref.~\onlinecite{Padro2002}) (better agreement 
is seen considering the calculation of minimum energy). $V_{zz}$ at the O2 site has 
discontinuous changes, but the EFG tensor is continuous, as will be shown later.

The coefficients in the $V_{zz}(P)$ expression obtained are 
shown in Table~\ref{tab:fitcoeff}, 
for all the atoms in the unit cell of all compounds considered.
At the $A$, $B$, and apical O1 atoms, the relation found in almost 
all the compounds considered here  is 
$\bigl( \begin{smallmatrix}
\Delta V_{xx} & 0 & 0 \\ 
0 & \Delta V_{yy} & 0 \\ 
0 & 0 & \Delta V_{zz} 
\end{smallmatrix} \bigr) = \Delta P^ 2
\times \bigl( \begin{smallmatrix} a_{xx} & 0 & 0 \\ 
0 & a_{yy} & 0 \\ 
0 & 0 & a_{zz} 
\end{smallmatrix} \bigr)$.
However, since these sites have $\eta=0$, the tensor is 
defined by only one independent parameter. $V_{xx}=V_{yy}=-1/2V_{zz}$, 
and the quadratic coefficients also follow the same symmetry
 $a_{xx}=a_{yy}=-1/2a_{zz}$. 
The equatorial oxygen sites (O2) do not have an $n$-fold rotation
 axis with $n\ge3$. This implies that the asymmetry parameter
 is not zero. In this case the coefficients of $P^2$ describing 
 the variation of the EFG tensor do not show such a simple relation. 
 
\begingroup
\begin{table*}\centering
\caption{\label{tab:fitcoeff}
Coefficients $a$ for the fits of the expression  
$V_{zz}(P)=V_{zz}^0+a\times P^2$. $V_{zz}^0$ in units 
of $10^{21}$ V\,m$^{-2}$, $a$ in units of $10^{22}$ V\,C$^{-2}\,$m$^{2}$, for the compounds with tetragonal and 
cubic perovskite type structures.}
\begin{threeparttable}
\begin{tabular}{c|c|c|c|c|c|c|c|c|c|c|c|c|c|c|c|c|c|c}
\hline\hline
&\multicolumn{2}{|c|}{BaTiO$_3$}&\multicolumn{2}{|c|}{KNbO$_3$}  & \multicolumn{2}{|c|}{PbTiO$_3$} &\multicolumn{2}{|c|}{BaZrO$_3$} & \multicolumn{2}{|c|}{CaTiO$_3$} &\multicolumn{2}{|c|}{PbZrO$_3$} & \multicolumn{2}{|c|}{SrTiO$_3$} & \multicolumn{2}{|c|}{NaNbO$_3$} & \multicolumn{2}{|c|}{LiNbO$_3$}\\

\hline
& $V_{zz}^0$ & $a$ &$V_{zz}^0$& $a$ &$V_{zz}^0$&\ \ \ \ $a$\ \ \ \ \ &$V_{zz}^0$& $a$ &$V_{zz}^0$& $a$ &$V_{zz}^0$& $a$ &$V_{zz}^0$& $a$ &$V_{zz}^0$& $a$ &$V_{zz}^0$& $a$ \\

 \hline
%%%%%%BaTiO3%%%%%%%%%KNbO3%%%%%%%%%%PbTiO3%%%%%%%%%%%%%%%%%%%%BaZrO3%%%%%%%%%%CaTiO3%%%%%%%%%%%%%%%%%%%%PbZrO3%%%%%%%%%%%SrTiO3%%%%%%%%%%NaNbO3%%%%%%%%%%LiNbO3%%%%%%%%
A & 0.16 & -1.079 & 0.14 & -0.128 & 5.28 & -1.787\tnote{a}\ &  0.00 & -0.945 &  0.00 & -0.186          & -0.01 & -1.911 & 0.00 & -0.551 & 0.00 & -0.077 & 0.00 & -0.012  \\
B & 0.78 & -0.949 & 3.60 & -1.046 & 5.58 & -0.562           &  0.00 & -0.757 &  0.01 & -0.724          & -0.02 & -0.985 & 0.01 & -0.834 & 0.00 & -0.163 & 0.00 & -0.311  \\
O1& 2.70 & -1.410 & 2.25 & -1.042 & 1.63 & -0.303           & -1.43 & -1.124 & -0.11 & -1.025          & -3.61 & -1.009 & 1.08 & -1.193 & 0.30 & -0.882 & 0.67 & -0.940  \\
O2& 2.32 &  0.288 & 1.37 &  0.625 & 1.18 & -0.461\tnote{b}\ & -1.41 &  0.300 & -0.10 &  0.349\tnote{c} & -3.59 &  0.232 & 1.11 &  0.253 & 0.32 &  0.389 & 0.69 &  0.363  \\
\hline\hline
\end{tabular}
     \begin{tablenotes}
       \item[a] An additional quartic term $b\times P^4$, with $b=3.626\times 10^{21}$ V\,C$^{-4}\,$m$^6$, is 
needed to get a satisfactory fit for this atom.
       \item[b] This coefficient fits a component that is $V_{zz}$ in the paraelectric phase, but is 
interchanged with other components in the distortion path. An additional quartic 
term $b\times P^4$, with $b=1.157\times 10^{21}$ V\,C$^{-4}\,$m$^6$, is 
needed to get a satisfactory fit for this atom.
       \item[c] This coefficient fits a component that is $V_{zz}$ in the paraelectric phase, but is 
interchanged with with other components in the distortion path. An additional quartic term $b\times P^4$, with 
$b=-4.643\times 10^{21}$ V\,C$^{-4}\,$m$^6$, is 
needed to get a satisfactory fit for this atom.
     \end{tablenotes}
\end{threeparttable}
\end{table*}
\endgroup

The variation of the EFG with displacements can be understood by 
considering 
a Taylor series expansion,
\begin{equation*}
V_{zz}=V_{zz}^0+\sum_i \frac{\partial V_{zz}}{\partial r_i}\delta r_i + 
\sum_i \frac{\partial^2 V_{zz}}{\partial r_i^2}\delta r_i^2 + \cdots , 
\end{equation*}
where $\delta r_i$ are small deviations of the atomic 
positions relative to the paraelectric structure. When the transition involves
 small displacements and for atoms where the EFG does not undergo 
large changes in the transition this expansion should converge rapidly. 
For sites without inversion symmetry, 
the  linear term should be  dominant, 
whereas for sites with inversion symmetry the linear term
 vanishes and the quadratic term in the expansion becomes 
relevant.~\cite{Dening1980}

The V$_{zz}$ component follows a quadratic variation for all 
the atoms in all the compounds studied, with only two exceptions,
 PbTiO$_3$ and CaTiO$_3$. 
At the O2 sites of PbTiO$_3$ and CaTiO$_3$ there are 
interchanges of tensor components which make
the description of EFG variations in terms of 
the $V_{zz}$ component inadequate.
At the Pb site of PbTiO$_3$ a small quartic 
term in the polarization is found necessary 
for a good fit of $V_{zz}(P)$ ($a\times P^2+b\times P^4$) 
(Terms with odd powers of $P$ are not allowed due to 
the inversion symmetry, in the paraelectric structure, of the sites 
involved in the distortion.)

\begin{figure}[tbp]
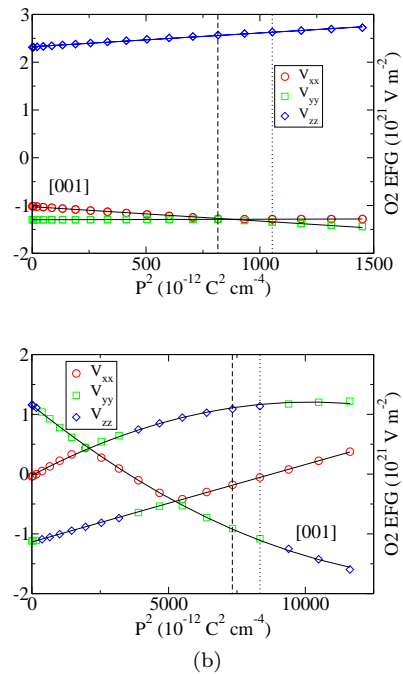

\begin{center} 
\subfigure[]{\label{efgbatio2}\includegraphics[width=0.6\linewidth]{fig3a.eps}}
\subfigure[]{\label{efgpbtio2}\includegraphics[width=0.6\linewidth]{fig3b.eps}}
\caption{(Color online) (a) BaTiO$_3$ in the tetragonal phase: $V_{zz}$, $V_{yy}$, and $V_{xx}$ components of the EFG tensor for 
the O2 atoms as a function of $P^2$. The lines are fits to the tensor components with regular variation. (b) 
PbTiO$_3$ in the tetragonal phase: $V_{zz}$, $V_{yy}$, and $V_{xx}$ components of 
the EFG for the O2 atoms as a function of $P^2$. The [001] direction curves are also 
indicated. Depending on the specific atom, the other two 
 curves correspond to either the [100] or [010] directions.}
\end{center}
\end{figure}

In the paraelectric structure of BaTiO$_3$ the principal component of 
the tensor $V_{zz}$ for Ba and Ti is directed along the $z$ axis.
For the O atoms, the EFG tensor is also aligned with the tetragonal  
crystalline axes, and $V_{zz}$ is directed to the neighboring Ti atoms.
 For BaTiO$_3$, with increasing ferroelectric distortion ($P$), 
in all distortions calculated, the direction of $V_{zz}$ remains 
the same for all atoms. The  $V_{xx}$ and $V_{yy}$ components also
 maintain their directions in this path, for the Ba, Ti and O1 atoms,
 along the $x$ and $y$ axes, respectively. For the O2 atoms, however,
  $V_{xx}$ and $V_{yy}$ do not always correspond to the same  
orientations. The three components of the EFG tensor for the O2 site 
in BaTiO$_3$ are shown in Fig.~\ref{efgbatio2}.  At a given distortion, 
due to the convention $|V_{yy}| \ge |V_{xx}|$ 
the regular curves followed by these components are interchanged. 
For distortion $\lambda\le 0.9$ the directions for $V_{xx}$, $V_{yy}$
 are $z$ and $x$ for O2 at (0.5,0,0.5); $z$ and $y$ for O2 
at (0,0.5,0.5). For higher $\lambda$ these directions are interchanged. 
Nevertheless, the quadratic behavior of EFG($P$) is maintained here 
and this interchange is ignored. Table~\ref{tab:fitcoeffO2} shows the coefficients of the quadratic
 terms for the variation of the EFG tensor components at each crystalline
 axis direction in BaTiO$_3$. Unlike in the case of $\eta=0$, the quadratic coefficients $a$ do 
 not have the same symmetry as the EFG tensor components.
 
\begin{table}\centering
\caption{\label{tab:fitcoeffO2}
Fit coefficients for the O2 atoms.
$V(P)=V^0+a\times P^2$. 
$V^0$ in units of $10^{21}$ V\,m$^{-2}$, $a$ in units 
of $10^{22}$ V\,m$^{-2}\,C^{-2}$.}
\begin{tabular}{lccc}
\hline\hline
Compound & EFG component & $V^0$ & $a$ \\ 
\hline
BaTiO$_3$ &$V_1$&-1.31&0.019\\
          &$V_2$&-1.03&-0.297\\
          &$V_3$&2.32&0.288\\
\hline\hline
\end{tabular}
\end{table}
The tensor component interchanges may only happen at the O2 
sites, where $\eta$ changes. For BaZrO$_3$, PbZrO$_{3}$, NaNbO$_{3}$, LiNbO$_{3}$ and SrTiO$_3$ there 
are no interchanges between tensor components, for KNbO$_3$ there is an 
interchange between $V_{xx}$ and $V_{yy}$, while for CaTiO$_{3}$ and PbTiO$_{3}$ there 
are exchanges between the three components.
For the sake of brevity, we only show here the O2 EFG tensor components 
for PbTiO$_3$ [Fig.~\ref{efgpbtio2}], where there 
are as many as five interchanges of one tensor component in the range of distortions 
considered. In particular, there is an interchange near the experimental distortion of 
which one should be aware in EFG measurements, since a kink measured in a $V_{zz}$ dependence 
could be a consequence of the convention that $V_{zz}$ is the largest EFG component in 
magnitude, instead of a phase transition.

Moreover, as already mentioned, to obtain a satisfactory fit to $V_{zz}(P)$ at the Pb 
site a small quartic term should also be considered [Fig.~\ref{pol2Vzzall_PbTi}]. $V_{zz}(P)$ follows 
approximately a linear behavior for larger values of $P^2$. At the O2 site two of the curves 
of components with constant direction also need quartic terms for a good fit [Fig.~\ref{efgpbtio2}]. 
The much larger displacements in this compound would indeed 
indicate that the $V_{zz}(\delta r)$ expansion does not converge as fast as in other cases.
In order to confirm that with larger displacements additional terms 
must be included in $V_{zz}(P)$, we performed the calculation of BaTiO$_3$ again, but this time 
we doubled the size of the distortions, allowing $0 \le \lambda \le 2.4$. The variation of the EFG ceases 
to be properly described by a single quadratic term at the O sites, and an additional term is needed, as expected.

For the other cases there is only an exception in the O2 sites of CaTiO$_3$ where a quartic term is also needed.
We point out that CaTiO$_3$ has the smaller unit cell (Table~\ref{tab:latt}), so the ions will be closer 
with the same $\lambda$ in comparison to the other compounds, which may be related to this exception.

In the pseudocubic cases the chosen distortions
are arbitrary. Nonetheless, we can conclude
from these cases that, for small distortions, a simple quadratic variation of 
$V_{zz}$ with polarization is seen in several 
different systems, apart from small deviations.

\subsection{Orthorhombic structure}\label{orthorhombic}
BaTiO$_3$ exhibits monoclinic, rombohedral, and orthorhombic phases at different temperatures. 
In order to see what are the differences in the EFG and $P$ with a change of structure, we have 
made a series of calculations in the orthorhombic phase. We also took one experimental 
measurement~\cite{Kwei1993} of the orthorhombic structure as the 
reference distortion ($\lambda =1$) and calculated from $\lambda=0$ to $\lambda = 1.2$ of this distortion 
keeping the lattice parameters ($a$, $b$, $c$) constant. The theoretical distortion of minimum energy 
is once again found to be close to 85\% of the experimental distortion. 

\begin{figure}[tbp]
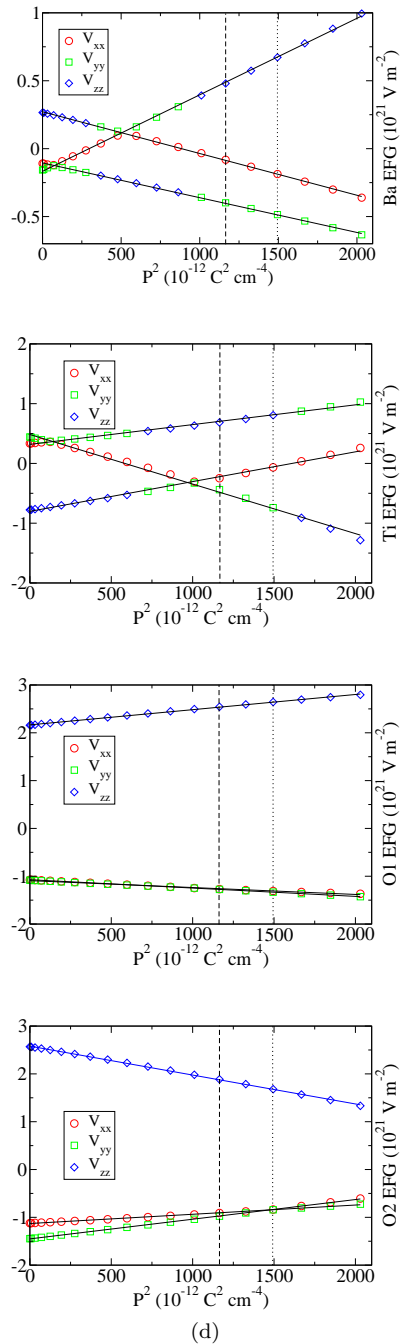

\begin{center} 
\subfigure[]{\includegraphics[width=0.6\linewidth]{fig4a.eps}}
\subfigure[]{\includegraphics[width=0.6\linewidth]{fig4b.eps}}
\subfigure[]{\includegraphics[width=0.6\linewidth]{fig4c.eps}}
\subfigure[]{\includegraphics[width=0.6\linewidth]{fig4d.eps}}
\caption{(Color online)\label{4fig} EFG tensor components in the principal coordinate system, 
of (a) Ba, (b) Ti, (c) O1 and (d) O2, as a function of $P^2$ for orthorhombic BaTiO$_3$. The lines are fits to the tensor components of regular variation.}
\end{center}
\end{figure}

The resolved components of the EFG tensors at the four inequivalent atoms against $P^2$ are displayed 
in Fig.~\ref{4fig}, and the coefficients resulting from the fits are listed in Table~\ref{tab:fitcoeffor}. 
In this structure there is not a single site with axial symmetry, and the relationship between 
EFG and $P$ will never be as simple as discussed above. 
There are also discontinuities in $V_{zz}$ at Ba and Ti sites due to the interchange of components. 
Nevertheless, ignoring these interchanges, the relation for each direction is purely quadratic in all cases. 
These results, like the previous ones, show that there is no physical meaning in the change of the 
principal axis definition (as $xx$, $yy$, or $zz$), and that the 
conventional assignment may obscure a simpler relation with crystal axes.

\begin{table}
\begin{center}
\caption{\label{tab:fitcoeffor}
Coefficients $a$ for the fits of the expression 
$V(P)=V^0+a\times P^2$ for BaTiO$_3$ in the orthorhombic 
phase, where $V_{1,2,3}$ is a component of the tensor with regular variation. 
$V_{1,2,3}$ in units 
of $10^{21}$ V\,m$^{-2}$, $a_{1,2,3}$ in units of $10^{22}$ V\,C$^{-2}$\,m$^{2}$.} 
\begin{tabular}{crrrr}
\hline\hline
 & Ba & Ti & O1 & O2 \\ 
  $V_1^0$  &  -0.17 & -0.80 & 2.17  & -1.13  \\
  $a_1$     & 0.563   & 0.498  & 0.318  & 0.193 \\
  $V_2^0$  & -0.11  &  0.48 & -1.07 & -1.45 \\
  $a_2$     & -0.256  & -0.829 & -0.174 & 0.411 \\
  $V_3^0$  & 0.27   & 0.32  & -1.09 & 2.58   \\
  $a_3$     & -0.307  & 0.331  & -0.144 & -0.605  \\
\hline\hline
\end{tabular}
\end{center}
\end{table}

\subsection{Correlations between EFG tensor components}\label{czjzek}

For the cases studied with tetragonal and cubic lattice parameters, $\eta\ne 0$ only for the O2 atoms. 
For $\eta=0$ the components of the EFG are trivially related, but when $\eta \ne 0$, the possible correlation 
between tensor parameters should be studied. The usual parametrization of $V_{zz}$ and $\eta$ assumes uncorrelated EFG components. 
Instead, correlations between the components of the 
tensors may be studied in a plot of one component against the other. However, just plotting $V_{zz}$ against $V_{xx}$ results 
in a distorted cobweb plot, and EFG tensor trajectories obtained by continuous variation of some parameter (temperature, 
distortion, pressure) may not be continuous (when $\eta=1$, $V_{zz}$ changes sign). Czjzek proposed a different system to 
eliminate these problems,~\cite{Czjzek1983} initially to deal with the analysis of amorphous systems. However, it is 
also suited to investigate the correlations between the tensor components.~\cite{Butz1996,Butz2010} This plot 
uses $-2V_{xx}$ as a function of $|2(2V_{zz}+V_{xx})/\sqrt{3}|$. With this linear combination of tensor components, the 
tractories are always continuous, and are straight lines if there is a linear dependence of $V_{zz}$ on $V_{xx}$. 

\begin{figure}[tbp]
\begin{center} 
\subfigure[]{\includegraphics[width=0.9\linewidth]{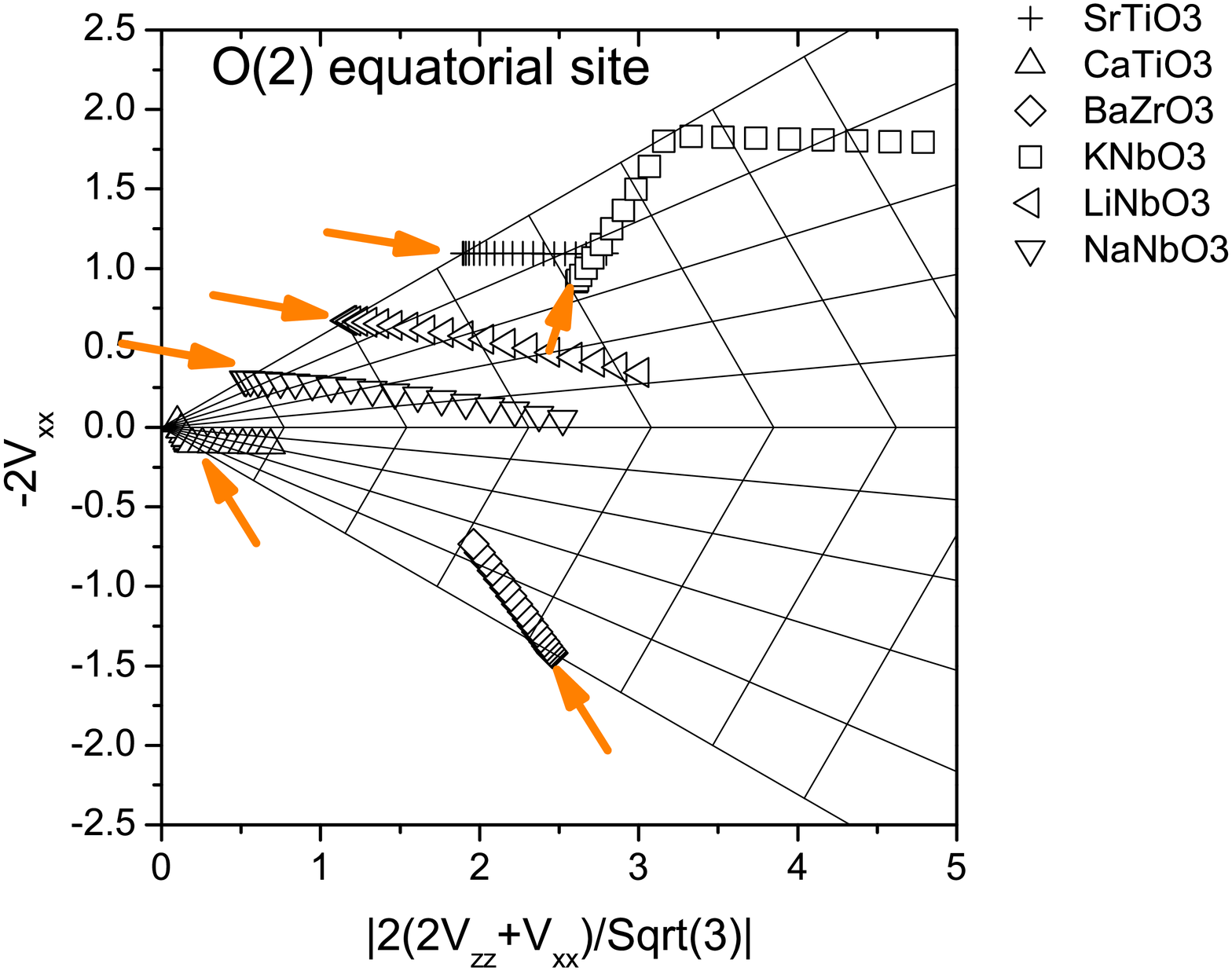}}
\subfigure[]{\includegraphics[width=0.9\linewidth]{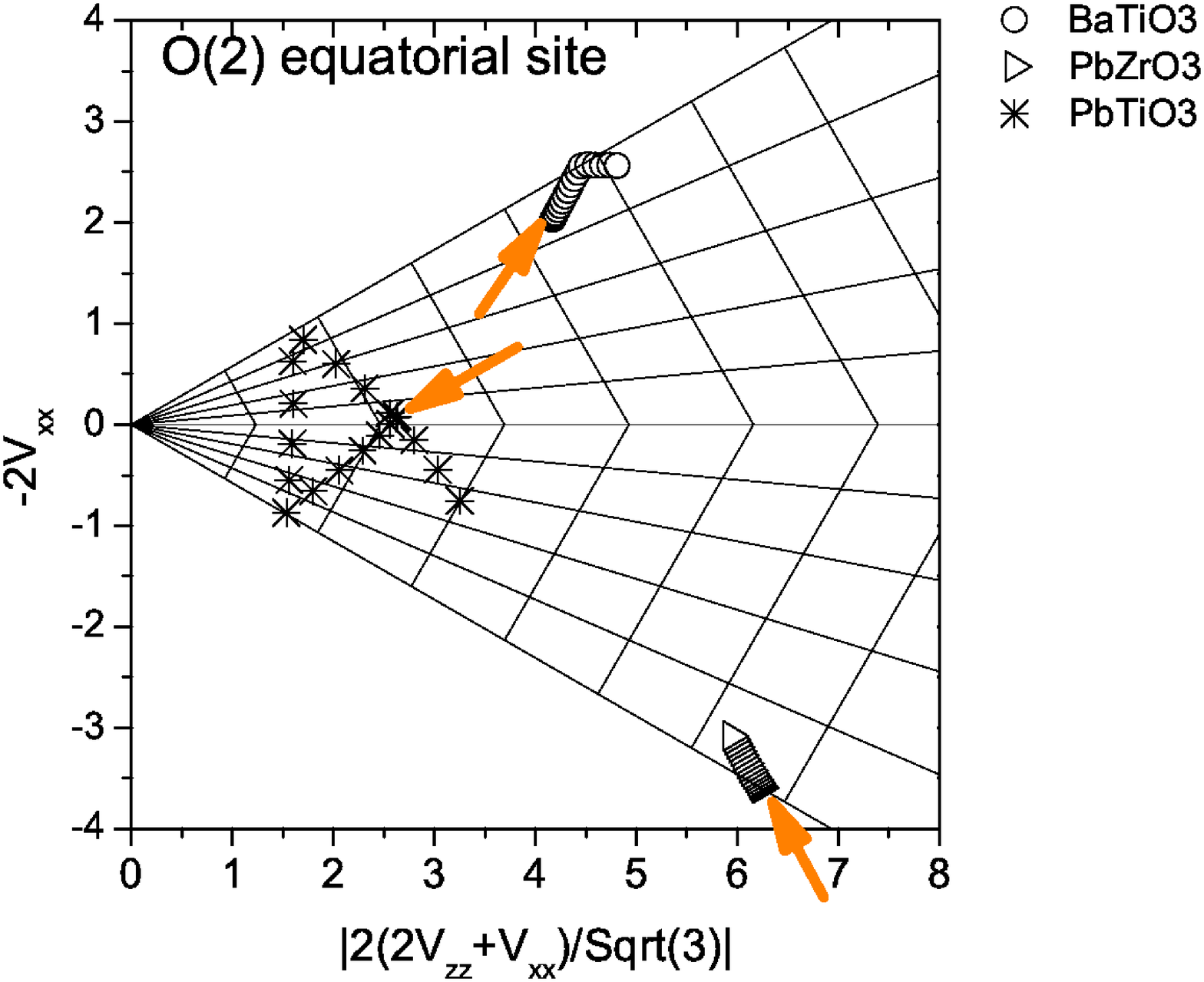}}
\caption{(Color online)\label{Czjzek} Cjzjek plots of the EFG tensor components at the equatorial O2 sites with $P$ (or distortion) 
as the implicit parameter in the perovskite materials studied, covering $0<\lambda<1.2$, in units of $10^{21}$ V\,m$^{-2}$. 
The arrows indicate the ``initial'' point and direction of the trajectories, corresponding to $P=0$.
Trajectories start at $\eta=0$ except for the tetragonal cases.}
\end{center}
\end{figure}

The trajectories in the Czjzek plot of all the tetragonal and pseudocubic cases are shown in Fig.~\ref{Czjzek} for the equatorial oxygen atoms, 
where $\eta$ changes and a non-trivial correlation may be present. In this plot, the 
lines of constant $\eta$ are the lines emerging from the origin: the boundary lines 
correspond to $\eta=0$ and the horizontal line corresponds to $\eta=1$. The herringbone lines correspond 
to constant $V_{zz}$. Reflections at the boundary of the plots, seen for example 
in BaTiO$_3$ and KNbO$_3$, are associated with the 
interchange of $V_{xx}$ and $V_{yy}$ components previously shown, and the crossing of the 
trajectory of the line $\eta=1$ corresponds 
to a change in sign and orientation of $V_{zz}$. The equatorial oxygen sites still have 
axial symmetry ($\eta=0$) in the paraelectric structure in the 
cubic structures, since there is four-fold rotation symmetry around the axes connecting the O to the B sites. 
Therefore, in these cases, the trajectories start at the boundary of the plot. 
For the tetragonal cases, this symmetry is lost and $\eta \ne 0$ even in the paraelectric phase: $\eta=0.12$ for 
BaTiO$_3$, $\eta = 0.33$ for KNbO$_3$ and $\eta=0.93$ for PbTiO$_3$. The upper (lower) wedge is for positive 
(negative) $V_{zz}$. The values of $V_{zz}$(O2) change significantly for the different compounds, even changing sign. 
KNbO$_3$, SrTiO$_3$, LiNbO$_3$ 
and NaNbO$_3$ always have positive values of $V_{zz}$, while the zirconates BaZrO$_3$ and PbZrO$_3$ have negative 
values of $V_{zz}$ in distortions considered. The EFG changes sign in the trajectories of CaTiO$_{3}$ and PbTiO$_3$. 
In the case of CaTiO$_3$ the trajectory overlaps with itself. It starts at negative 
EFG and $\eta=0$, then it goes to positive EFGs, in the direction of the arrow, it is 
reflected and comes back the same way, changing sign again, and finally it is reflected to 
the horizontal path. In general the positive EFGs increase in absolute value with distortion, while 
the negative ones decrease. Exceptions in this respect are parts of the trajectories in PbTiO$_3$ and CaTiO$_3$. 

The lengths of the trajectories in this plot are presented in Table~\ref{lengths}. 
\begin{table}
\begin{center}
\caption{\label{lengths} Approximate values for the lengths of the EFG(O2) trajectories 
in the Cjzjek plots, relative to the length of BaTiO$_3$, and their asymptotic asymmetry parameter ($\eta_{\infty}$).} 
\begin{tabular}{crr}
\hline\hline
 Compound & length   &$\eta_{\infty}$ \\ 
  BaTiO$_3$  &   1   &     0.92     \\
  SrTiO$_3$  & 0.99  &   1     \\
  CaTiO$_3$  & 0.85  &  1  \\
  BaZrO$_3$  & 0.88  & 0.81 \\
  LiNbO$_3$  & 1.91  & 0.63  \\
  NaNbO$_3$  & 2.02  & 0.77 \\
  PbZrO$_3$  & 0.59  & 0.92 \\
  KNbO$_3$   & 2.87  & 1  \\
  PbTiO$_3$  & 5.76  & 0.46 \\
\hline\hline
\end{tabular}
\end{center}
\end{table}
The lengths are markedly larger for PbTiO$_3$, KNbO$_3$, LiNbO$_3$, and NaNbO$_3$. 
The cases of PbTiO$_3$ and KNbO$_3$ correspond to different fractional distortions in 
comparison to the other cases. However, all the other cases have fractional distortions 
equal to BaTiO$_3$. In this respect, it is interesting to remark that both 
niobates (LiNbO$_3$ and NaNbO$_3$) have trajectories with approximately twice 
the length of the others, indicating that the local charge at the O2 atoms is more 
sensitive to displacements when Nb is the $B$ site. 

From these plots it is also of 
interest to determine the ``asymptotic'' asymmetry parameter ($\eta_{\infty}$),~\cite{Butz2010} related to 
the slope of the trajectories after reflections at the boundaries, corresponding to the 
limiting value of $\eta$ for a given trajectory. These values 
are presented in Table~\ref{lengths}. It can be seen that $\eta_{\infty}$ is large in 
most cases, with a smaller value for PbTiO$_3$. It should be related to the different features of the distortion, due to the fact that 
this compound has ferroelectricity driven by the 
lone pair of $s$ electrons at the Pb ions, while ferroelectricity in other 
compounds is related to the $d^0$ configuration at the $B$ site.

If the trajectories in this plot are straight lines, this shows that the EFG tensor components 
are linearly related. This is usually the case, except for LiNbO$_3$ and NaNbO$_3$, where the 
trajectory is approximately straight for low values of $P$, but becomes curved for higher values. 
The linearity in almost all cases means that the whole tensor, when considering the distortions 
that give rise to ferroelectric polarization, can be described by a \emph{single} parameter. 
This might have important implications on the way experimental data should be analyzed, since 
the components are not independent, and the combination of all information 
will be more useful than each of them taken separately. 
The study of the EFG dependence on temperature (or on other variables) usually done 
by separating $V_{zz}$ and $\eta$, can be performed by identifying the global single parameter.

\section{Relation between EFG and $P$ in different materials}\label{Arelation}

\begin{figure}[tpb]
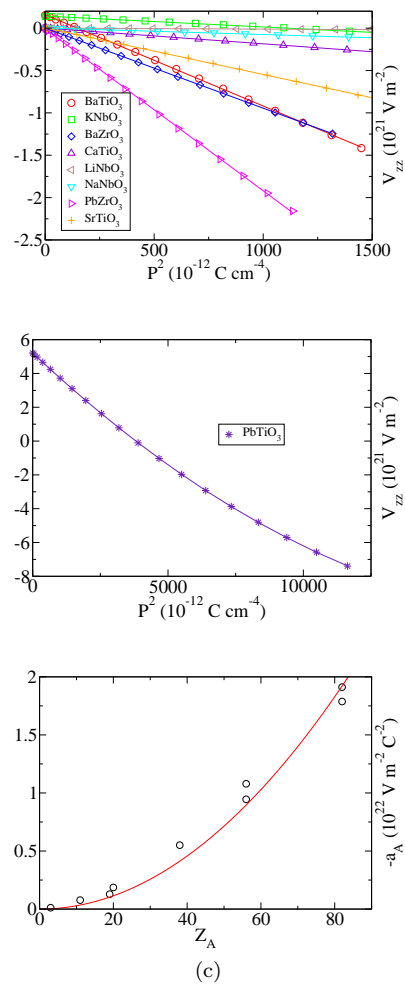

\begin{center} 
\subfigure[]{\label{a_Za1}\includegraphics[width=0.6\linewidth]{fig6a.eps}}
\subfigure[]{\label{a_Za2}\includegraphics[width=0.6\linewidth]{fig6b.eps}}
\subfigure[]{\label{a_Za3}\includegraphics[width=0.6\linewidth]{fig6c.eps}}
\caption{(Color online)\label{a_ZA} (a) $V_{zz}$ at the A site as a function of $P^2$, showing the 
different coefficients (slopes of straight lines). (b) $V_{zz}$ as a function of $P^2$ for PbTiO$_3$.
(c) Coefficients of quadratic term of $V_{zz}(P)$, at the A site, for a series of materials with the perovskite structure, 
as a function of atomic number $Z_A$. The line is a quadratic fit.}
\end{center}
\end{figure}

In the following we analyze the relations found in different materials. 
The quadratic coefficients $a$ in the $V_{zz}(P)$ expression follow an interesting trend with atomic 
number at the $A$ sites ($Z_A$), shown in Fig.~\ref{a_ZA}. Figures~\ref{a_Za1} and~\ref{a_Za2} 
show $V_{zz}$ at the $A$ site for the different cubic and tetragonal compounds. PbTiO$_3$ introduces larger variations 
for $P^2$ and $V_{zz}$ which may  follow from the larger displacements in the experimental distortion, 
related, as previously discussed, to the different nature of the ferroelectricity.

$V_{zz}$ is $0$ for the paraelectric cubic cases, while 
for the paraelectric tetragonal structures V$_{zz}^A$ has small nonzero values, except 
PbTiO$_3$. The slopes of the trajectories correspond to the coefficients of the quadratic term, 
which are shown in Fig.~\ref{a_Za3}. The $A$ sites with a larger number of electrons show a larger 
magnitude of $a$, that follows a quadratic dependence on $Z_A$.
There are two pairs of compounds 
with the same $Z_A$, but different metal $B$ sites, namely BaTiO$_3$/BaZrO$_3$, and PbTiO$_3$/PbZrO$_3$. 
The same trend is still qualitatively followed although small changes in the values of the coefficient 
are seen, and the changes do not follow a pattern: $a$ is higher for $B=\ $Ti than $B=\ $Pb in the cases with $A=\ $Ba, 
but the opposite happens when $A=\ $Pb. 
We did not find other trends for the other coefficients, or for $V_{zz}^0$, as a 
function of the atomic number of the different sites.

\section{Conclusions}\label{Conclusions}

In summary, our calculations provide an \textit{ab initio} support for previous
 observations of a quadratic dependence of the EFG tensor
versus $P$ in ferroelectric materials with the perovskite structure and other distorted structures. 
An exception to this rule is only seen in PbTiO$_3$, at the Pb and equatorial O sites, and in CaTiO$_3$ at 
the equatorial sites, where an additional quartic term dependence is observed. Moreover, for most cases the 
local symmetry of the atom means that the EFG tensor is axially symmetric and all the components are 
trivially related to each other. For the equatorial oxygen atoms this is not the case. Nevertheless, the components of the 
EFG tensor are also quadratically related to $P$ and there is a linear correlation among EFG tensor components for most cases. 
The relation between $P$ and EFGs follows a trend with the atomic number of the $A$ site, qualitatively a quadratic variation. 

The EFG, working as a local analog of $P$, has the added advantage of higher spatial resolution 
and short time scales, which allows local probing of nanoscale phenomena at specific lattice sites. 
The different types of lattice sites, including defects, can be discriminated by their different EFGs. 
Its critical behavior in phase transitions may be analyzed, and the order of the transitions can be established 
with a high degree of detail. In hysteresis loops, it can act as a 
static measurement of the electric polarization in individual domains, not limited by depolarization effects and is 
much more sensitive to polarization reversal. It is also suited to probe phase coexistence or inhomogeneous 
polarization states with atomic selectivity, well beyond the reach of conventional polarization measurements.  
Moreover, piezoelectric force microscopy is usually restricted to studies near the surface of samples, while 
EFG studies can be performed in bulk or at the surface of materials, by using diffusion, evaporation or 
implantation techniques to add the probe atoms in the environments to study.

Therefore, we hope this work will stimulate more EFG studies. 
However, one limitation of our results should be mentioned. In this paper we have analyzed the 
variation of the EFG with the scaling of the ferroelectric 
distortions as a whole. However, in experiments, the parameters, such as temperature, in 
general produce variations involving, for example, additional distortion modes, lattice vibrations, and lattice expansions. 
The variation of each independent atomic 
displacement may have a complex variation with changing temperature.  Therefore, the relation investigated here 
can not be directly used to infer polarization variations from EFGs or vice-versa, except in 
well understood cases. Further work to improve this limitation could involve the study 
of different structural changes or temperature effects.

\section*{Acknowledgments}

This work has been supported by 
the AQUIFER (Aquila Initiative for Ferroics) research program, sponsored 
by the International Center for Materials Research (ICMR) at UCSB, 
and research projects PTDC/FIS/105416/2008 and CERN/FP/116320/2010. 
J.\ N.\ Gon\c{c}alves acknowledges FCT Grant No. SFRH/BD/42194/2007. 
The theoretical research at CNR-SPIN has received funding by the European 
Community's Seventh Framework Programme 
FP7/2007-2013 under Grant No. 203523-BISMUTH.
A.~S. thanks P.~Barone for comments on the manuscript. 
Computational support by CASPUR Supercomputing center in Rome is acknowledged.

\end{document}